\documentclass[conference]{IEEEtran}
\IEEEoverridecommandlockouts

\usepackage{graphicx} 
\usepackage{subcaption}

\usepackage{cite}
\usepackage{amsmath,amssymb,amsfonts}
\usepackage{algorithmic}
\usepackage{graphicx}
\usepackage{textcomp}
\usepackage[dvipsnames]{xcolor}
\usepackage{xcolor}
\usepackage{url}
\usepackage{hyperref}
\hypersetup{
     colorlinks=false,
     }

\def\BibTeX{{\rm B\kern-.05em{\sc i\kern-.025em b}\kern-.08em
    T\kern-.1667em\lower.7ex\hbox{E}\kern-.125emX}}
\begin{document}

\title{HamilToniQ: An Open-Source Benchmark Toolkit for Quantum Computers
\thanks{The first two authors contributed equally to this work.}
}

\author{\IEEEauthorblockN{Xiaotian Xu}
\IEEEauthorblockA{\textit{QuEST} \\
\textit{Imperial College London}\\
London, United Kingdom \\
xiaotian.xu19@imperial.ac.uk}
\and
\IEEEauthorblockN{ Kuan-Cheng Chen}
\IEEEauthorblockA{\textit{QuEST} \\
\textit{Imperial College London}\\
London, United Kingdom \\
kuan-cheng.chen17@imperial.ac.uk}
\and
\IEEEauthorblockN{Robert Wille}
\IEEEauthorblockA{\textit{Chair for Design Automation} \\
\textit{Technical University of Munich}\\
Munich, Germany \\
robert.wille@tum.de}

}

\maketitle

\begin{abstract}
In this paper, we introduce HamilToniQ, an open-source, and application-oriented benchmarking toolkit for the comprehensive evaluation of Quantum Processing Units (QPUs). Designed to navigate the complexities of quantum computations, HamilToniQ incorporates a methodological framework assessing QPU types, topologies, and multi-QPU systems. The toolkit facilitates the evaluation of QPUs' performance through multiple steps including quantum circuit compilation and quantum error mitigation (QEM), integrating strategies that are unique to each stage. HamilToniQ's standardized score, H-Score, quantifies the fidelity and reliability of QPUs, providing a multidimensional perspective of QPU performance. With a focus on the Quantum Approximate Optimization Algorithm (QAOA), the toolkit enables direct, comparable analysis of QPUs, enhancing transparency and equity in benchmarking. Demonstrated in this paper, HamilToniQ has been validated on various IBM QPUs, affirming its effectiveness and robustness. Overall, HamilToniQ significantly contributes to the advancement of the quantum computing field by offering precise and equitable benchmarking metrics.
\end{abstract}


\section{Introduction}
In the context of rapid developments in quantum computer hardware and software, an increasing number of quantum algorithms have been proven to possess quantum supremacy\cite{boixo2018characterizing,arute2019quantum}, with many practical applications demonstrating the foresight of quantum computing. During this progress, quantum computers continue to be developed using various hardware architectures, such as superconducting circuits\cite{devoret2013superconducting}, ion traps\cite{kielpinski2002architecture}, cold atoms\cite{weiss2017quantum}, and photonic quantum computing\cite{yu2023universal}, among others. On the software front, there are platforms like IBM’s Qiskit\cite{wille2019ibm,Qiskit2023}, Google’s Cirq\cite{Cirq2021}, and Xanadu's PennyLane\cite{bergholm2018pennylane} available. This diversity in the ecosystem offers researchers a range of choices; however, there currently lacks a clear set of criteria to guide researchers on which quantum hardware or corresponding software is more suitable for specific application scenarios. Especially as quantum computers move towards integration with high-performance computing (HPC) ecosystems, the creation, validation, and implementation of benchmarks for quantum computing become a foundational aspect of quantum computer architecture evolving alongside HPC towards Quantum-HPC. 

In the late 1970s, the computing boom led to the development of LINPACK and SPEC for benchmarking supercomputers and workstations\cite{dongarra2003linpack,henning2006spec}. The rapid advancement of personal computers gave rise to the PARSEC benchmark suite, which was designed for Chip-Multiprocessors. Similarly, the swift development of applications in machine learning led to the creation of MLPerf for benchmarking performance across various ML models. Likewise, in the era of rapid development in quantum computing, it is imperative to establish a new benchmarking system tailored for quantum computer architecture, including quantum computing techniques\cite{kordzanganeh2023benchmarking
  }, device topology\cite{mckinney2023co}, compilation strategy\cite{ferrari2023modular} and quantum-error-mitigation protocol\cite{strikis2021learning,cai2023quantum}, for testing, design, and optimization.

Thus, researchers and developers have already proposed tools for benchmarking the general performance of Quantum Processing Units (QPUs). The Quantum Volume, proposed by IBM Quantum\cite{jurcevic2021demonstration,bishop2017quantum}, serves as a metric to evaluate the number of qubits in a system and the complexity of operations that can be executed. It accounts for various factors including gate and measurement errors, device cross-talk, connectivity, and compiler efficiency. Additionally, Sandia National Laboratories has introduced the circuit mirroring technique, which ingeniously converts any quantum program into a series of ``mirror circuits". These circuits maintain the complexity of the original program while producing predictable outcomes\cite{proctor2022measuring,blume2020volumetric}. Furthermore, Infleqtion has developed the SupermarQ Suite, an application-oriented suite for evaluating various algorithms on QPUs\cite{tomesh2022supermarq}, including GHZ-state and the Variational Quantum Eigensolver (VQE). A research group from the Technical University of Munich has also developed MQTBench, an open-source tool that facilitates QPU benchmarking by providing over 70,000 benchmark circuits, thereby enhancing the transparency and fairness of QPU performance comparisons\cite{quetschlich2023mqt}.

Our study integrates application-oriented principles and open-source accessibility, focusing on evaluating existing QPUs, compilation strategies, quantum-error mitigation techniques, and distributed quantum systems using the Quantum Approximation Optimization Algorithm (QAOA). We introduce HamilToniQ—a comprehensive benchmarking toolkit designed to facilitate comparative analysis across various QPUs. Unlike conventional benchmarking software, HamilToniQ operates on a standardized user-level platform with uniform assessment metrics and incorporates a novel scoring system with in-depth metrics specific to QAOA algorithms. This approach aims to provide researchers with more effective and equitable benchmarking scores (H-Score). 

The paper is structured as follows: Section \ref{ch: design} provides a detailed exposition of the architecture and technical specifications. Performance metrics and benchmarking results for a range of real QPUs are presented in Section \ref{Sec: results}, where the efficacy and robustness of HamilToniQ are also evaluated, demonstrating its value and capability within the quantum computing field. 



\section{Background}
\subsection{QAOA and Q-Matrix}
HamilToniQ benchmarks a QPU based on the performance of the optimization algorithm running on it, typically employing the QAOA algorithm.

QAOA is a hybrid quantum-classical algorithm designed for solving combinatorial optimization problems involving $n$ variables \cite{farhi_quantum_2014}. Its primary aim is to find a set of binary values $x \in \{0, 1\}^n$ that minimize a quadratic objective function, expressed as:
\begin{align}
    x^TQx
\end{align}
Each of the binary values indicates a distinct choice within the problem domain, with $1$ representing ``Yes'' and $0$ representing ``No''. The matrix $Q$, a real symmetric matrix within $\mathbb{R}^{n \times n}$, encodes the optimization problem, known as the ``Q matrix". Q matrix encodes the combinatorial optimization problem, by putting the correlation of a pair of variables on the off-diagonal terms, and the variance of each variable on the diagonal terms. One example of the problem set is shown in Fig. \ref{fig:qaoa1}. The QAOA algorithm has demonstrated significant advantages in the fields of portfolio optimization\cite{brandhofer2022benchmarking}, routing\cite{azad2022solving}, and scheduling\cite{pelofske2023quantum} problems.

A parameterized quantum circuit (PQC), known as the ansatz, is used to generate a quantum state, represented by a vector $\Tilde{x}$. The binary values are the measurement results of this quantum state on the computational basis. A classical optimization algorithm iteratively alters the trainable parameters in this ansatz to minimize the cost function $\Tilde{x}^TQ\Tilde{x}$, thereby guiding the quantum state towards the correct result.

QAOA distinguishes itself from other variational quantum algorithms through its universal ansatz. It begins with a uniform superposition of all possible states on the computational basis. This initial state is then manipulated by alternating layers of phase and mixer operators, which are embedded with trainable parameters. The Q matrix is encoded into the phase operators. The modularity of QAOA allows the number of layers—or the circuit's depth—to be adjusted, accommodating different levels of problem complexity.  

\subsection{Quantum Circuit Compilation Flow}

In the advancement of quantum computing, the theoretical constructs of quantum algorithms require translation into executable protocols that align with the specific constraints of physical QPUs. Initially designed within a hardware-agnostic framework, these algorithms must be accurately adapted to operate within the practical confines of real-world quantum systems. This transition from theory to application is paramount in leveraging quantum computation's potential.

\begin{figure}[htpb] 
    \centering 

    \begin{subfigure}[b]{0.17\textwidth} 
        \includegraphics[width=\textwidth]{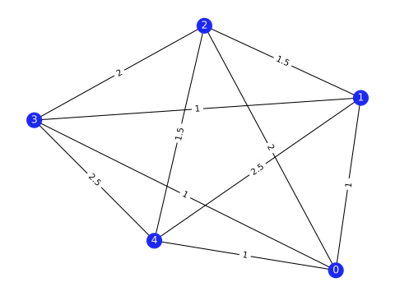}
        \caption{An example of QAOA problem set.}
        \label{fig:qaoa1}
    \end{subfigure}
    \hfill 
    \begin{subfigure}[b]{0.28\textwidth}
        \includegraphics[width=\textwidth]{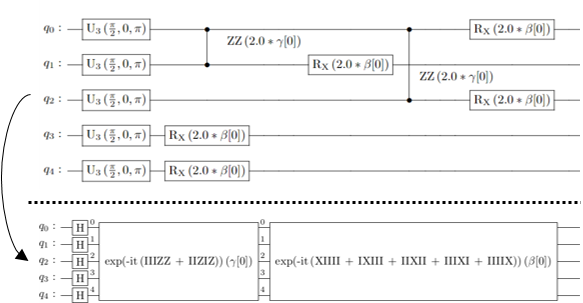}
        \caption{Target-independent Circuit}
        \label{fig:qaoa2}
    \end{subfigure}

    \vspace{0.5cm} 

    \begin{subfigure}[b]{0.45\textwidth} 
        \includegraphics[width=\textwidth]{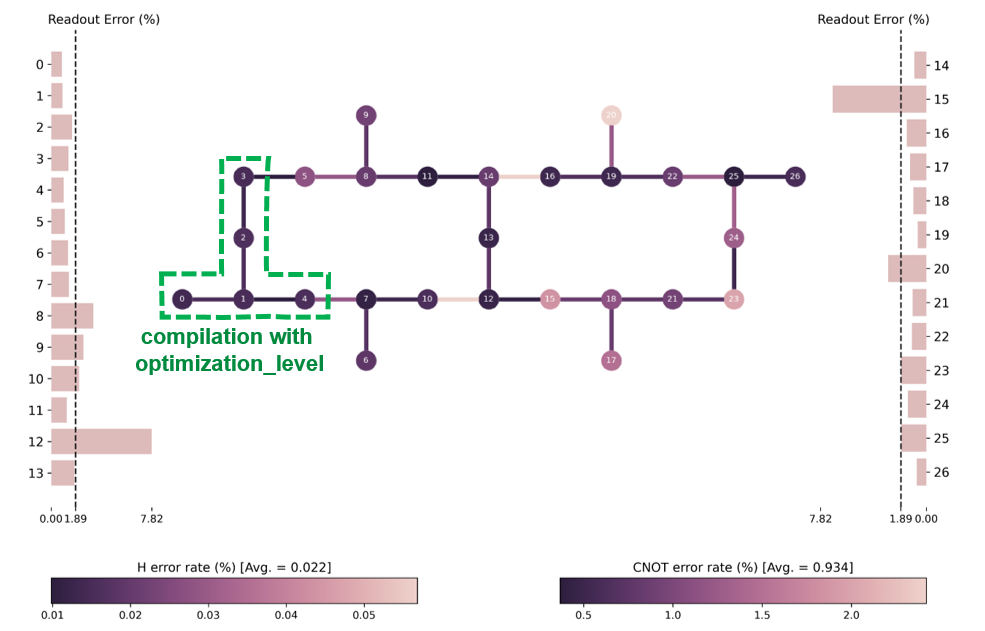}
        \caption{Compilation of target-dependent QAOA circuit}
        \label{fig:qaoa3}
    \end{subfigure}

    \caption{Compilation Process of a QAOA Algorithm.} 
    \label{fig:qaoacompilation}
\end{figure}

The QAOA serves as an example of this translation process. It necessitates the transformation of classical problem sets into quantum domains through the construction of a Pauli Hamiltonian, as depicted in Fig. \ref{fig:qaoacompilation}(a). The encoding of this Hamiltonian is critical; it must be structured such that its lowest cost corresponds to the correct answers, represented by the maximum number of interconnecting edges between disparate node groups. Following the establishment of the Hamiltonian, the QAOA ansatz is materialized in the form of quantum circuits, as shown in Fig. \ref{fig:qaoacompilation}(b). The circuit is then subject to an optimized compilation strategy, tailored to leverage the most stable qubits available on the QPU. Notably, in systems like those developed by IBM, this entails adherence to the instruction set architecture (ISA), ensuring that both the ansatz and the observables conform to the specifications of the chosen quantum system. Utilizing tools such as the \texttt{qiskit.transpiler} and its \texttt{optimization\_level} parameter, the compilation strategy is refined to optimize circuit performance for the backend's ISA. Successfully navigating this compilation phase is critical; it allows for the precise mapping of the quantum algorithm to the QPU's most stable qubits, as shown in Fig. \ref{fig:qaoacompilation}(c), which is essential for maintaining benchmarking workflow consistency and obtaining dependable results. The circuit compilation process can also benefit from reinforcement learning approaches, as demonstrated by the MQT Predictor\cite{quetschlich2023mqt}.

\subsection{Quantum Error Mitigation and Other Error-Suppression Protocols}
In the NISQ (Noisy Intermediate-Scale Quantum) era, the noise inherent in quantum computers significantly impacts the results of quantum computing\cite{preskill2018quantum}. Noise affects the accuracy of our algorithmic results and the assessment of system performance. Thus, there are many quantum error mitigation techniques have been developed\cite{cai2023quantum}. In the benchmarking of QPUs, the influence of noise on algorithms is indirectly incorporated into our scoring benchmarks, which are based on the noise’s impact on the accuracy of algorithmic outcomes. Consequently, our evaluation of QPU performance includes a subcomponent for quantum error mitigation. Protocols such as Twirled Readout Error eXtinction (T-Rex)\cite{van2022model} and Zero Noise Extrapolation (ZNE)\cite{giurgica2020digital}, Probabilistic Error Cancellation (PEC)\cite{mari2021extending}, as well as dynamic decoupling (DD) techniques, are available on platforms like the IBM Quantum Hub to enhance the accuracy of algorithms on NISQ devices, with these improvements validated by multiple studies\cite{kim2023evidence,prest2023quantum, chen2023short}. In HamilToniQ, we propose that when setting up algorithms, quantum error mitigation protocols or other protocols that enhance algorithm accuracy can also be integrated into the benchmarking process. This approach not only makes the benchmarking process more reflective of the conditions under which researchers operate quantum computing algorithms but also allows HamilToniQ to directly benchmark and validate these protocols, thereby expanding the scenarios in which the software can be utilized.

\subsection{Quantum Software Stack}
\begin{figure}[ht]
    \centering
    \includegraphics[width=0.9\linewidth]{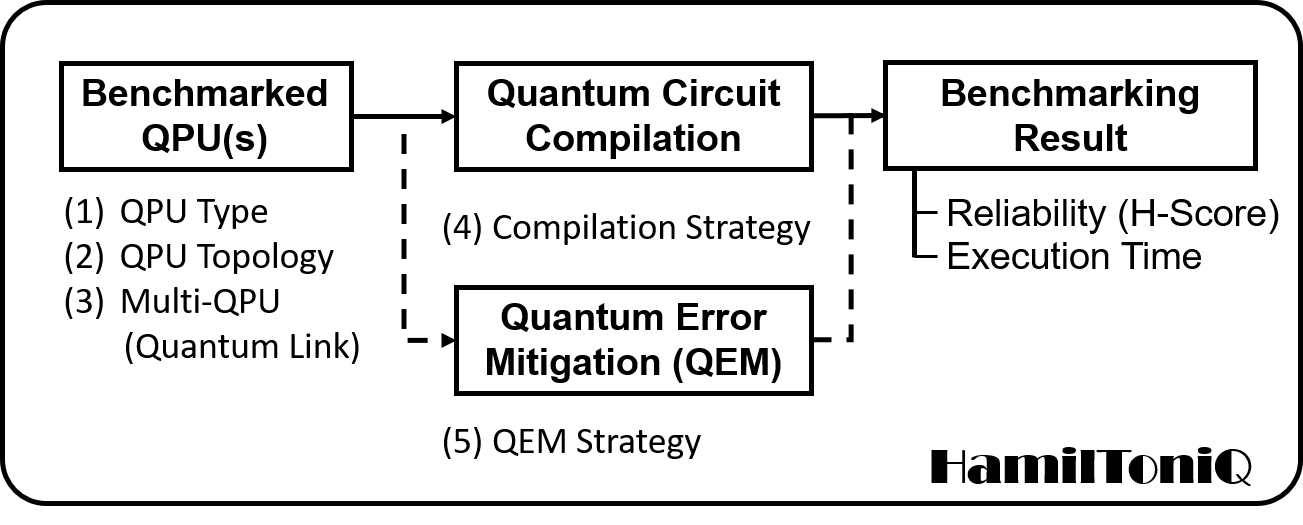}
    \caption{Flowchart of HamilToniQ Benchmarking Software Workflow: Demonstrating the integration of QPU characteristics, compilation and error mitigation strategies, and their effects on benchmarking outcomes for enhanced reliability (H-Score) and execution time. }
    \label{fig: hamiltoniq}
\end{figure}

The HamilToniQ's benchmarking workflow, shown in Fig. \ref{fig: hamiltoniq}, commences with the characterization of QPUs, where each QPU is classified according to its type, topology, and multi-QPU system,. This initial step ensures a tailored approach to the benchmarking process, considering the unique attributes of each QPU. Subsequently, the process engages in quantum circuit compilation, employing a specific strategy designed to optimize the execution of quantum circuits on the identified QPUs. Integral to the workflow is Quantum Error Mitigation (QEM), which strategically addresses computational noise and errors that could affect the fidelity of the quantum processes. The culmination of this rigorous workflow is the benchmarking result, which quantifies the performance of the QPU in terms of reliability—represented by the H-Score and Execution Time. These metrics provide a quantitative and objective measure of the QPU's performance, reflecting the effectiveness of the benchmarking process implemented by HamilToniQ. Additionally, the H-score can help manage computational resources in a Quantum-HPC system.

\section{Design}
\label{ch: design}

In this section, the architecture and technical details of the HamilToniQ are discussed. Additionally, details of the code are mentioned, specifically focusing on the components of the \verb|hamiltoniq| package and the \verb|Toniq| class, which is frequently utilized in benchmarking processes.

The core concept behind HamilToniQ is to assign higher H-Scores to QPUs that demonstrate greater accuracy on specified optimization tasks. This accuracy, defined as the probability of identifying the correct answer, is influenced by a variety of factors, including the noise, the complexity of the task, and the algorithm's configuration. The final H-Score of a QPU relies on statistical methods, guaranteeing an unbiased evaluation of processor performance. QAOA is employed as the representative of optimization algorithms in HamilToniQ. 

\begin{figure}[ht]
    \centering
    \includegraphics[width=0.9\linewidth]{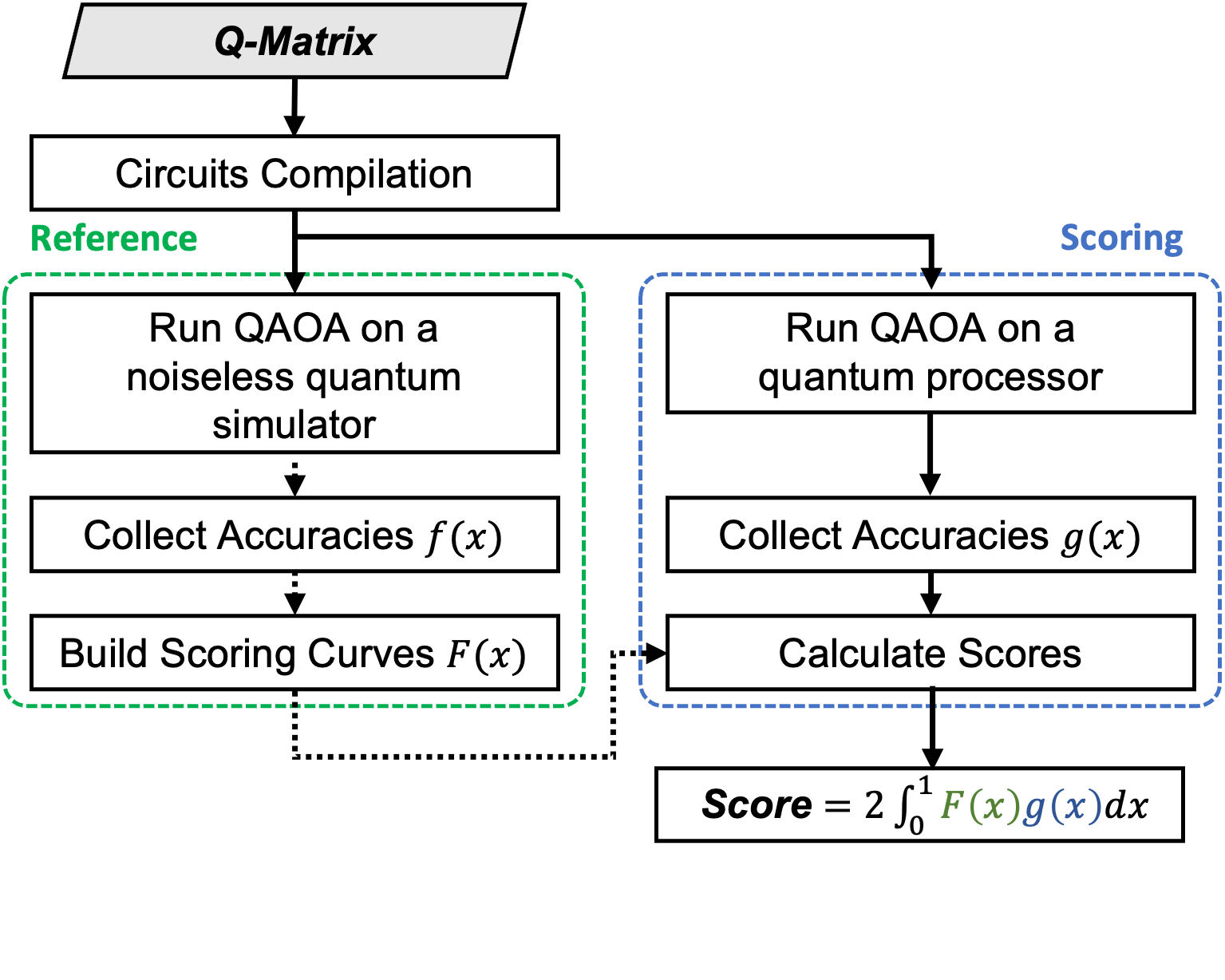}
    \caption{This flow chart illustrates the standard benchmarking procedures of a QPU using HamilToniQ. Users can use the Q matrices and scoring curves provided within \texttt{hamiltoniq.instances}. Alternatively, the \texttt{Toniq.get\_reference} function in \texttt{hamiltoniq.benchmark} can calculate the scoring curves of users' own Q matrices. The performance depends on the accuracy of solving QAOA problems using each QPU.}
    \label{fig: flow}
\end{figure}

HamilToniQ primarily comprises two components: the reference part, also known as ground truth, and the scoring part, as depicted in Fig. \ref{fig: flow}. The reference part, which is optional, utilizes a noiseless simulator to find the scoring curve. Users will only need this part when they are benchmarking on their own Q matrices. In the scoring part, the Quantum Approximate Optimization Algorithm (QAOA) is executed for a certain number of times, and the scoring curve is used to determine the score of each iteration based on its accuracy. The final H-Score is computed as the average of all individual scores.

\subsection{Instance}

Utilizing HamilToniQ necessitates the specification of a particular instance, especially for QAOA, where this instance is represented by a Q matrix. The Q matrix is a symmetric matrix that encodes a quadratic unconstrained binary optimization (QUBO) problem. This is achieved by allocating each off-diagonal element to represent the correlation between a pair of variables, while each diagonal element indicates the variance of an individual variable. Q matrices of different complexity \cite{brandhofer_benchmarking_2022} result in different probabilities of arriving at the correct results, and consequently, different H-scores. Hence, maintaining consistency in the Q matrix throughout the benchmarking process is essential.

Once the instance is decided, the brute-force method is then used to explore the solution space and find the lowest cost, along with the corresponding binary string and decimal number. This binary string is regarded as the ``correct answer" in the benchmarking procedure.

Users have the flexibility to use their own Q matrices. This can be accomplished by finding the scoring curve with \verb|Toniq.get_reference| and assigning the Q matrices to the \verb|Q| argument in \verb|Toniq|'s methods \verb|simulator_run| or \verb|processor_run|. By doing so, users can know which QPU is likely to yield more accurate results for their own instances. 

By default, the built-in Q matrices are used. In HamilToniQ, a series of Q matrices corresponding to 3, 4, 5, and 6 qubits are provided. These matrices were randomly generated, but specially selected to avoid an excessive number of results converging towards $0$ or $1$, thereby ensuring that the instances are neither too difficult nor too simplistic \cite{brandhofer_benchmarking_2022} (also ensuring that most of optimization results see a monotonically increasing curve, as discussed in Sec. \ref{Ch: performance}). By using the built-in Q matrices, standardized H-Scores are obtained, allowing for comparison with the H-Scores of other users. The Q matrices, as well as their lowest cost (\verb|energy|), binary string (\verb|bin_state|), and decimal state (\verb|dec_state|) are stored in \verb|hamiltoniq.instances|. The results in Sec. \ref{ch: result1} are obtained based on the built-in Q matrices.

\subsection{Building the Reference using a Quantum Simulator}
\label{Ch: reference}

\begin{figure}
    \centering
    \includegraphics[width=1\linewidth]{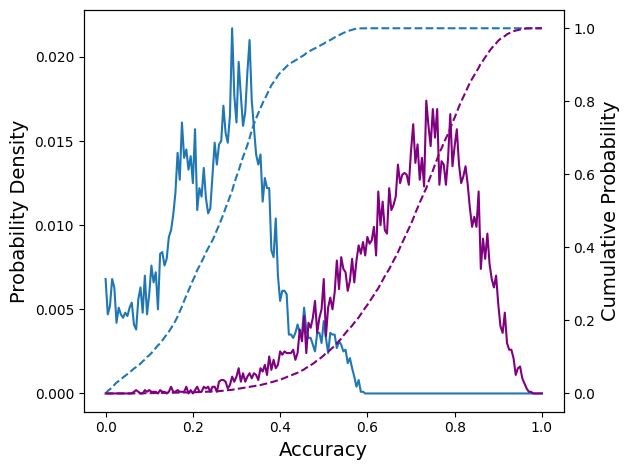}
    \caption{An example illustrates the distributions of accuracy (solid lines) and their cumulative sum (dashed lines). The cumulative sum is also the scoring curve mentioned in Sec. \ref{Ch: reference}. Two configurations of the QAOA were used: one with a single layer (blue lines) and another with 9 layers (purple lines). It is apparent that QAOA with 9 layers performs better than the other one, as more results have higher accuracy. This conclusion shows the shift of the peak of accuracy distribution towards the right in the figure. The results were obtained from the \texttt{qubit\_3} instance given in \texttt{hamiltoniq.matrices}.}
    \label{fig: pdf_scoring}
\end{figure}

In the reference part, the QAOA optimization is initially executed $N$ times on an ideal, noiseless simulator. In each iteration, the algorithm tries to find the lowest cost, subsequently returning the corresponding quantum state as the result. The accuracy is evaluated by squaring the inner product of the returned quantum state and the correct answer, thereby representing the probability of accurately measuring the correct answer from the returned quantum state. With a sufficiently large $N$, we can consider the distribution of accuracy as a probability density function (PDF) $f(x)$, where $x$ is the independent variable, accuracy, ranging from $0$ to $1$. An example of the distribution of accuracy is shown as the solid lines in Fig. \ref{fig: pdf_scoring}. The cumulative distribution function of the distribution of accuracy, denoted as $F(x)$, is the scoring curve (shown as the dashed lines in Fig. \ref{fig: pdf_scoring}). It is also defined as the integral of $f(x)$ over the range from $0$ to $x$:
\begin{align}
    F(x) = \int_0^x f(x')dx'
    \label{eqn: scoring}
\end{align}

In the code, the distribution of accuracy is determined by computing the histogram of all accuracy. Subsequently, the cumulative sum of this histogram and a 1-D interpolation are utilized to generate the scoring curve.

In the setting of HamilToniQ, the number of repetitions $N$ is set to $10,000$. This number is selected to provide a robust statistical basis for accurately assessing performance without requiring excessive computational time. Additionally, we can utilize the cuQuantum SDK with multi-GPU processing to accelerate the reference part\cite{bayraktar2023cuquantum, lykov2023fast, chen2024quantum}.

\subsection{Question: What is a Better Performance?}
\label{Ch: performance}

Before introducing the scoring procedure, we first need to define the criteria by which the QPUs are evaluated. In optimization problems, accuracy is the thing with the most attention, making it suitable for the criteria. Instead of simply using average accuracy for the statistical final H-Score, which diminishes the effect of an individual result, the idea of the integral of the product between the PDF and a monotonically non-decreasing function is used in HamilToniQ.

Consider two individual accuracy $a$ and $b$, with $a>b$. The PDF of those two results are two Dirac delta functions $p_1(x) = \delta(x-a)$ and $p_2(x) = \delta(x-b)$. In this scenario, $\delta(x-b)$ can be regarded as $\delta(x-a)$ with a shift to the left along the x-axis (x-axis denoting the accuracy). This shift can be represented as the variance between two delta functions:
\begin{align}
    \psi_{a\to b}(x) = \delta(x-b) - \delta(x-a)
\end{align}

Consider a monotonically non-decreasing function $h(x)$, the integral between this function and the variance is:
\begin{align}
    \int_0^1 \phi_{a\to b}(x) h(x)dx = h(b) - h(a)
    \label{eqn: single score}
\end{align}
This value must be non-positive. If both $a$ and $b$ are at the monotonically increasing part of $h(x)$, this value then must be negative. We would say, relative to PDF $p_2$, $p_1$ shows a better performance. This conclusion aligns with the fact that $p_1$ corresponds to a higher accuracy.

In HanilToniQ, the criteria of better performance is defined as the collective effect of Eqn. \ref{eqn: single score}. Let's consider two groups of results, each has $N$ pieces of results in it. These two groups can be expressed in the form of PDFs $P_1(x)$ and $P_2(x)$. If we still pick up result pairs and consider the shift, the average variance of the shift effect is:
\begin{align}
    \psi(x) &= \frac{1}{N} \sum_i^N \left[\delta(x-b_i) - \delta(x-a_i)\right]\\
    &=P_2(x) - P_1(x)
\end{align}

Again, consider a monotonically non-decreasing function $h(x)$, the integral between this function and the variance is:
\begin{align}
    \int_0^1 \phi(x) h(x)dx = \int_0^1 \phi(x) P_2(x)dx - \int_0^1 \phi(x) P_1(x)dx
    \label{eqn: collective score}
\end{align}
If this value is non-positive, we would say, relative to $P_2$, $P_1$ shows better performance. If the monotonically non-decreasing function is properly chosen, most of the results see the increasing part of $h(x)$ and this value is even negative. In HamilToniQ, we choose the scoring function obtained in Eqn. \ref{eqn: scoring}. This function not only puts most of the accuracy at the monotonically increasing part, but also processes self-normalization property, which will be mentioned in Sec. \ref{Ch: scoring}. This section's idea will be further explained in Sec. \ref{Ch: compare} for its impact on scoring.

\begin{figure*}[ht]
    \centering
    \begin{subfigure}[b]{0.32\textwidth}
        \includegraphics[width=\textwidth]{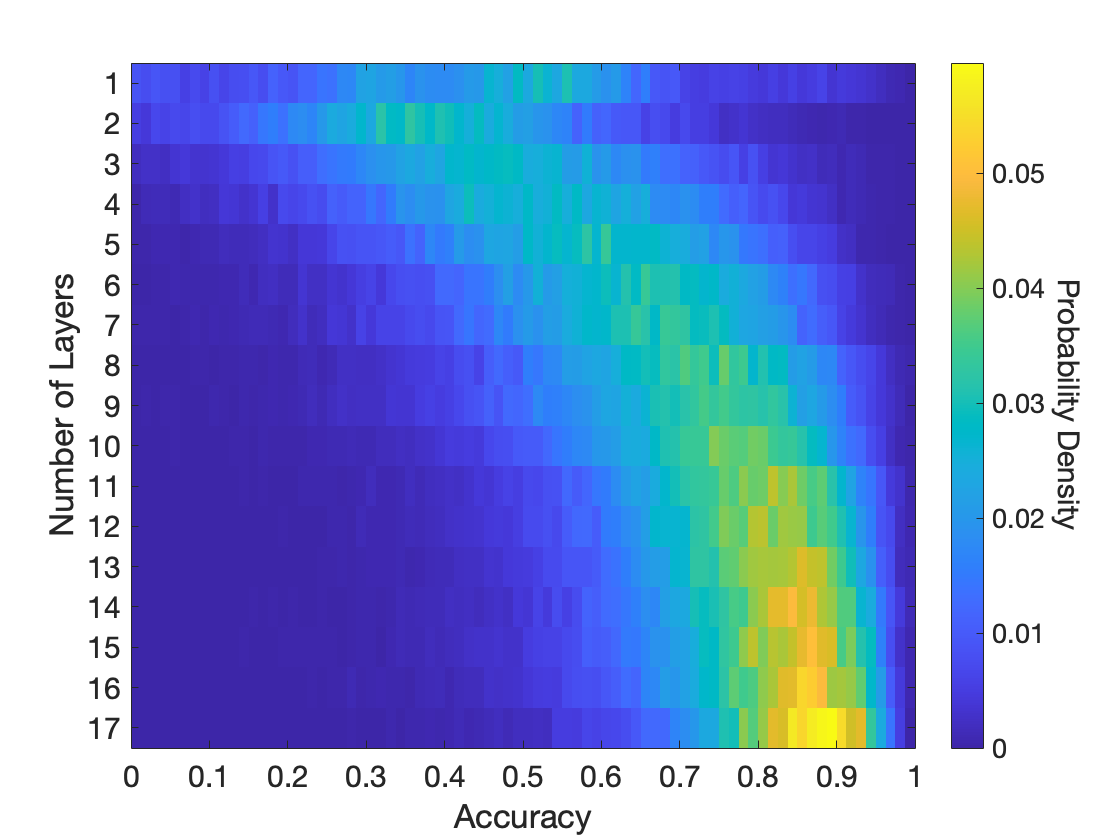}
        \caption{A noiseless simulator using 3 qubits}
        \label{fig:accuracy_simu_3_qubits}
    \end{subfigure}
    \hfill 
    \begin{subfigure}[b]{0.32\textwidth}
        \includegraphics[width=\textwidth]{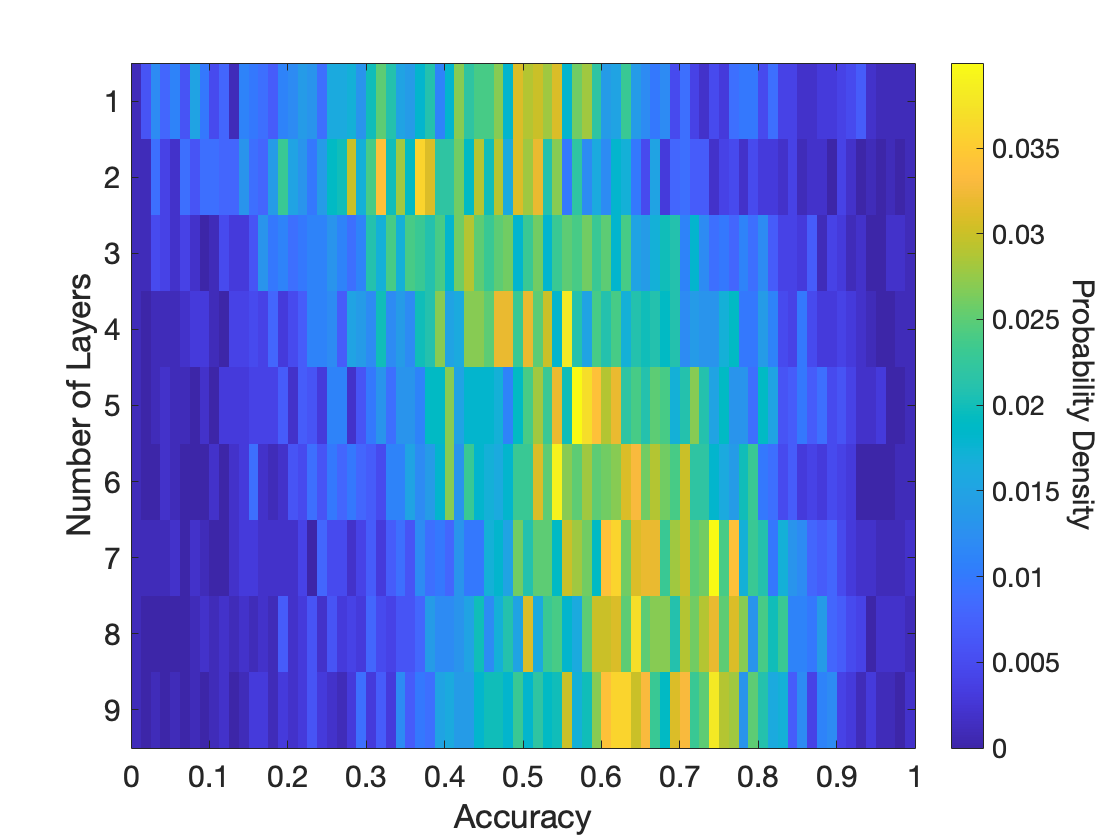}
        \caption{\textit{ibm\_lagos} using 3 qubits}
        \label{fig:accuracy_lagos_3_qubits}
    \end{subfigure}
    \hfill 
    \begin{subfigure}[b]{0.32\textwidth}
        \includegraphics[width=\textwidth]{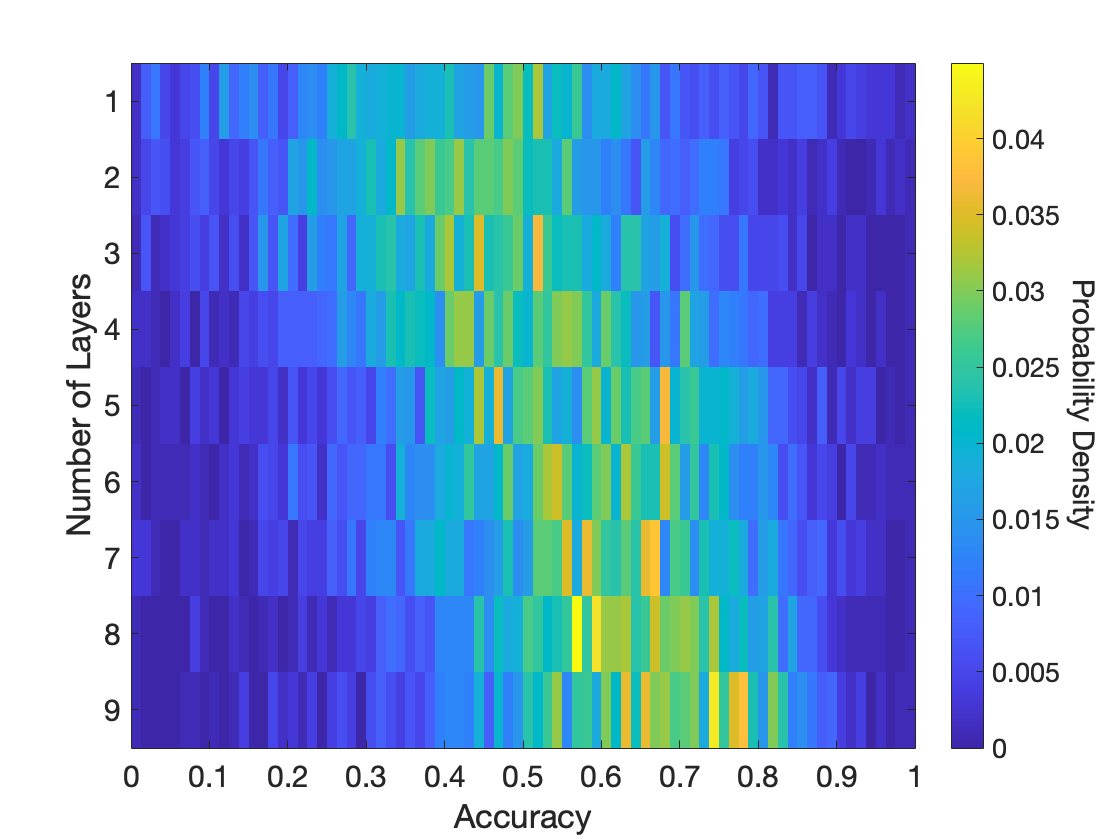}
        \caption{\textit{ibm\_cairo} using 3 qubits}
        \label{fig:accuracy_cairo_3_qubits}
    \end{subfigure}
    \caption{The accuracy distributions are illustrated for (a) a noiseless simulator using 3 qubits, (b) \textit{ibm\_lagos} using 3 qubits, and (c) \textit{ibm\_cairo} using 3 qubits. The vertical axis indicates the number of layers and the color represents the probability density. The brighter the color, the higher the probability density. In (a), the distribution for 1 layer and 9 layers correspond to the blue solid line and purple solid line in Fig. \ref{fig: pdf_scoring}, respectively. These results were obtained on \texttt{qubit\_3} instances given in \texttt{hamiltoniq.instances}.}
    \label{fig: accuracy}
\end{figure*}

\subsection{Scoring}
\label{Ch: scoring}

In the scoring part, QAOA is executed for $M$ times. And in each iteration, the accuracy is taken as the result. Let's denote the accuracy obtained in the $i$-th iteration as $X_i$. The score of this iteration is determined as $F(X_i)$, where $F(x)$ is the scoring function obtained from Eqn. \ref{eqn: scoring}. By averaging across all scores, the final H-Score $C$ of a QPU can be expressed as:
\begin{align}
    C = \frac{2}{M}\sum_i^M F(X_i)
    \label{eqn: sum}
\end{align}
note that a factor of 2 is placed on the numerator to accomplish the normalization (Eqn. \ref{eqn: normalization}).

When the number of iterations \(M\) is sufficiently large, it becomes feasible to consider the distribution of outcomes as a PDF, denoted by \(g(x)\), where \(x\) represents the accuracy, ranging from 0 to 1. In this scenario, the H-Score becomes the expected value of the scoring curve \(F(x)\) over the distribution \(g(x)\), which is written in the form of an integral:
\begin{align}
    C = 2 \int_0^1 F(x)g(x)\,dx \label{eqn: int}
\end{align}
also, a factor of 2 is used for normalization.

HamilToniQ uses Eqn. \ref{eqn: sum} to calculate the H-Score, while the other expression, Eqn. \ref{eqn: int}, demonstrates the self-normalization of the H-Score. When benchmarking a noiseless QPU, PDF $f(x)$ should be returned, which is the same distribution function observed on a noiseless simulator. Then, according to Eqn. \ref{eqn: int}, the H-Score of a noiseless QPU is:
\begin{align}
    C_\text{nl} = 2\int_0^1 F(x)f(x)dx
\end{align}
where the subscript ``nl" denotes ``noiseless". Since $F(x)$ is the primitive function of $f(x)$, it is natural to apply the integration by parts, which yields:
\begin{equation}
\begin{aligned}
    &2\int_0^1 F(x)f(x)dx\\
    =&2\left[F^2(x)\right]^1_0 - 2\int_0^1F(x)f(x)dx  
    \label{eqn: normalization}
\end{aligned}
\end{equation}

The scoring curve \(F(x)\) is a monotonically non-decreasing function that changes from \(0\) to \(1\). Therefore, the value of the term \(2\left[F(x)^2\right]_0^1\) is 2. After combining the same term on both sides, we find that the final H-Score of a noiseless QPU is always \(1\).
\begin{align}
    C_\text{nl} = 1
    \label{Eqn: normalization2}
\end{align}

However, if the performance of a QPU exceeds that of a noiseless simulator, the H-Score can be higher than \(1\). An extreme example occurs when, in every iteration, the optimization algorithm has a \(100\%\) probability of finding the correct answer, resulting in all scores being equal to \(1\) and the final H-Score reaching \(2\). This number represents the mathematical upper limit of H-Scores.

Besides, it is recommended to benchmark the fake backends, to anticipate the performance of real QPUs

In HamilToniQ, the number of repetitions $M$ is $1000$.
.
\subsection{Extra bit: Comparing two PDFs}
\label{Ch: compare}

The concept of self-normalization also further clarifies the scoring mechanism. The PDF of a QPU can be interpreted as the PDF observed on a noiseless simulator with a variation \(\psi(x)\) added, expressed as \(\psi(x) = f(x) - g(x)\). Given that both distribution functions, \(g(x)\) and \(f(x)\), are normalized over the interval \([0, 1]\), the difference \(\psi(x)\) is a function with the following relationship:
\begin{align}
    \int_0^1 \psi(x) = 0
\end{align}

Accordingly, the final H-Score corresponding to $g(x)$ is written as the sum of two integrals:
\begin{align}
    C = 2\int_0^1 F(x)f(x)dx + 2\int_0^1 F(x)\psi(x)dx
    \label{eqn: score with variance}
\end{align}
where the first term, as shown in Eqn. \ref{eqn: normalization}, evaluates to 1. Since $F(x)$ is always a monotonically non-decreasing function, the other term characterizes the overall impact of noise and errors on the QPU, as discussed in Sec. \ref{Ch: performance}. The value of $C$ is larger than $1$ if the QPU performs better than the noiseless simulator, and negative otherwise.

Upon extending the PDFs $f(x)$ and $g(x)$ to two generalized PDFs, Eqn. \ref{eqn: score with variance} provides a criterion for comparison them: If the integration $2 \int_0^1 F(x)g(x)dx$ is larger than $1$, $g(x)$ indicates a better performance; Conversely, a value less than 1 suggests worse performance. This comparison is conducted by calling the function \verb|Toniq.compare_distr|. With this function, HamilToniQ can also compare the results from the same QPU with a different number of layers, as well as compare the performance of different algorithms.

\section{Results}
\label{Sec: results}

\subsection{Proof of Concept: Distribution of Accuracy Between Noiseless and Noisy Conditions}

In this section, we demonstrate how to analyze the PDF, shown in Fig. \ref{fig: pdf_scoring}, for different numbers of layers, as illustrated in the flowchart from Fig.~\ref{fig: flow}. For proof of concept, Fig.~\ref{fig: accuracy} illustrates the accuracy distributions obtained from noiseless simulators and two noisy QPUs, \textit{ibm\_lagos} and \textit{ibm\_cairo}, across varying numbers of layers. In our results shown in Fig.~\ref{fig: accuracy}(a), for the noiseless simulator, accuracy generally improves with the addition of more layers. However, this trend is less pronounced with noisy QPUs, due to increased quantum gate error rates and the more significant impact of decoherence time associated with additional layers. Based on these results, we use the formula described in Eq.~\ref{eqn: sum} to calculate the H-Score.

\subsection{Benchmarking QPUs}
\label{ch: result1}
\begin{figure}[ht]
    \centering
    \includegraphics[width=1\linewidth]{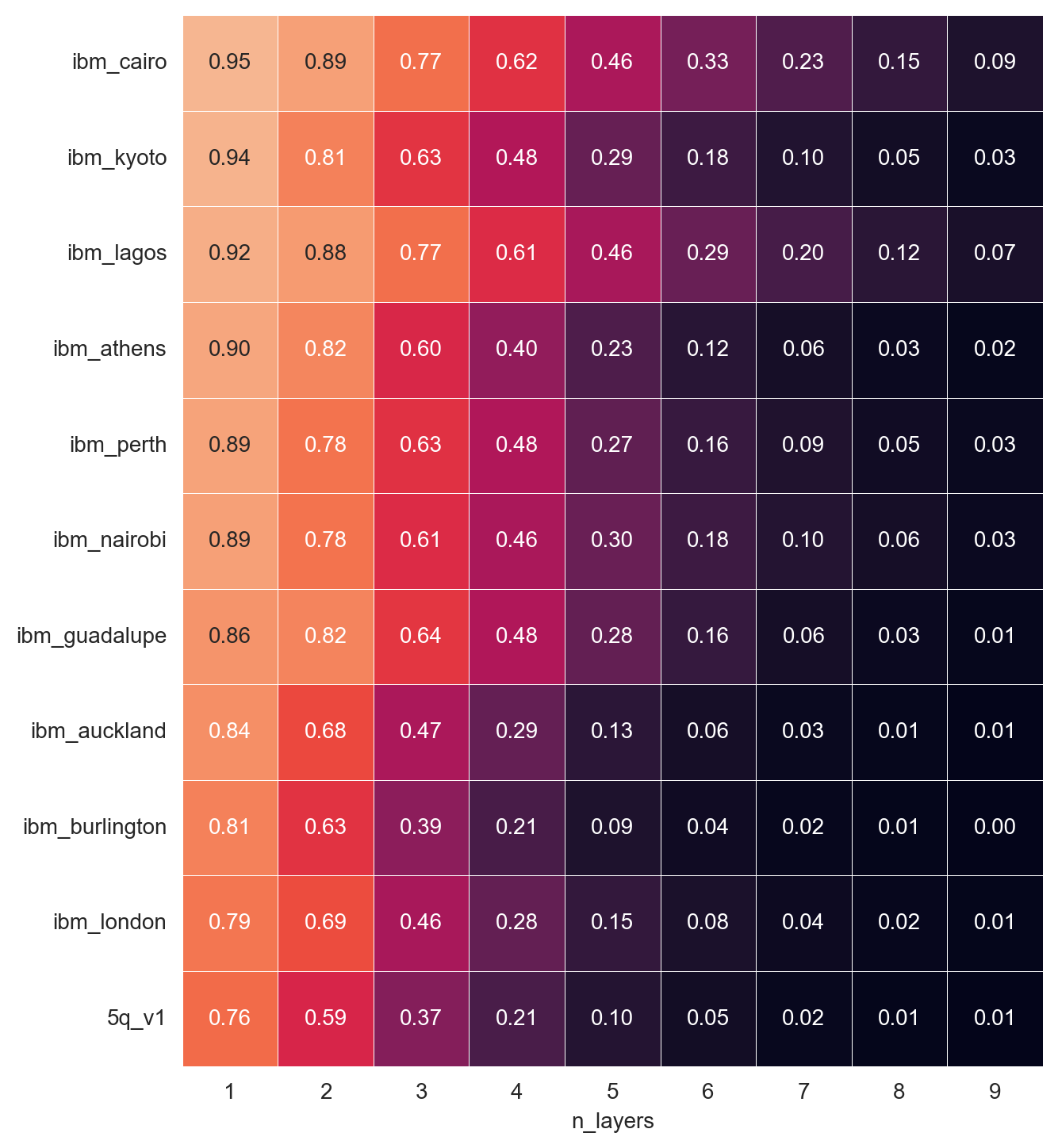}
    \caption{Benchmarking results on IBM QPUs with 3 qubits. The chart is sorted by the H-Scores with \texttt{n\_layers}=1. }
    \label{fig: result 3 qubits}
\end{figure}

\begin{figure}[ht]
    \centering
    \includegraphics[width=1\linewidth]{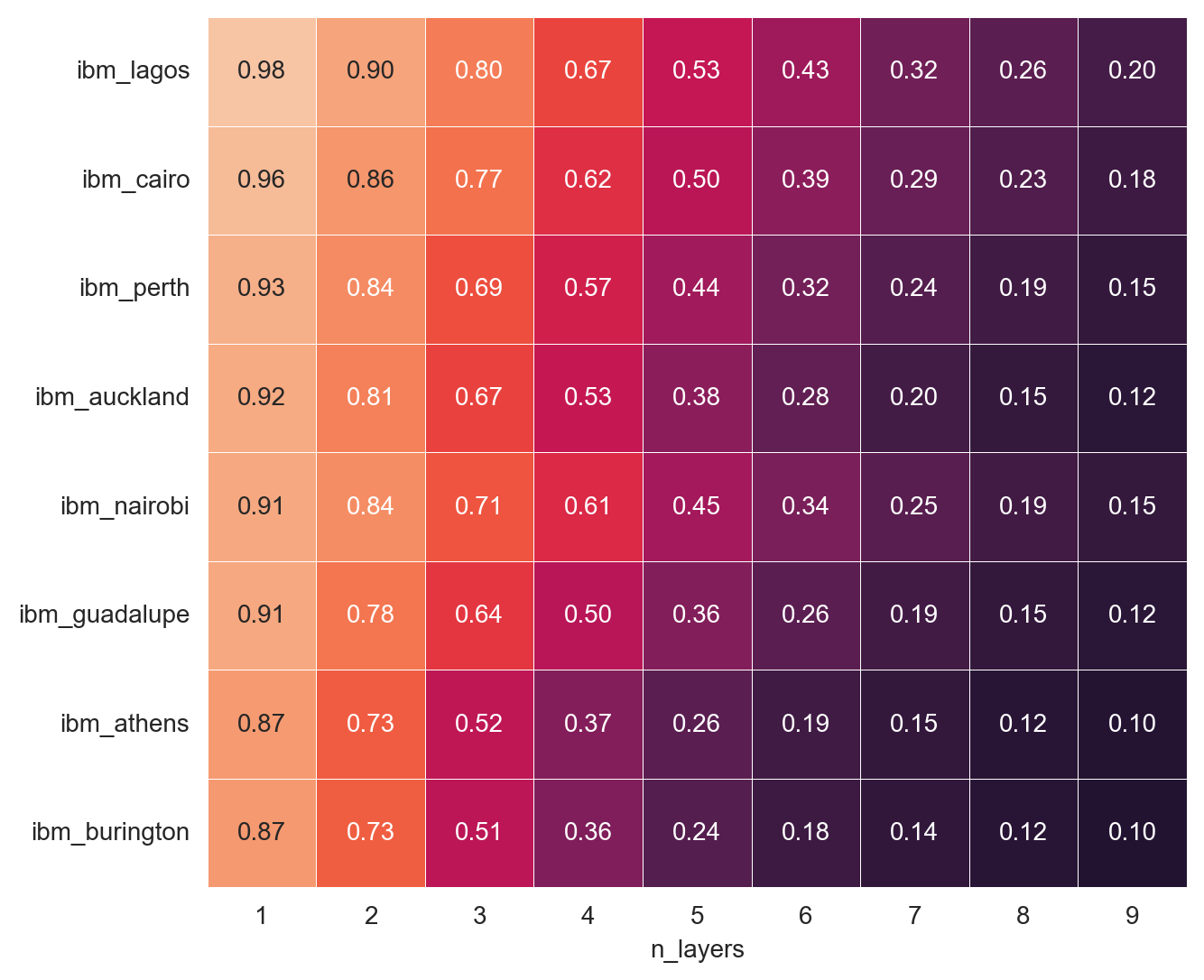}
    \caption{Benchmarking results on several IBM QPUs with 4 qubits. The chart is sorted by the H-Scores with \texttt{n\_layers}=1.}
    \label{fig: result 4 qubits}
\end{figure}

\begin{figure}[ht]
    \centering
    \includegraphics[width=1\linewidth]{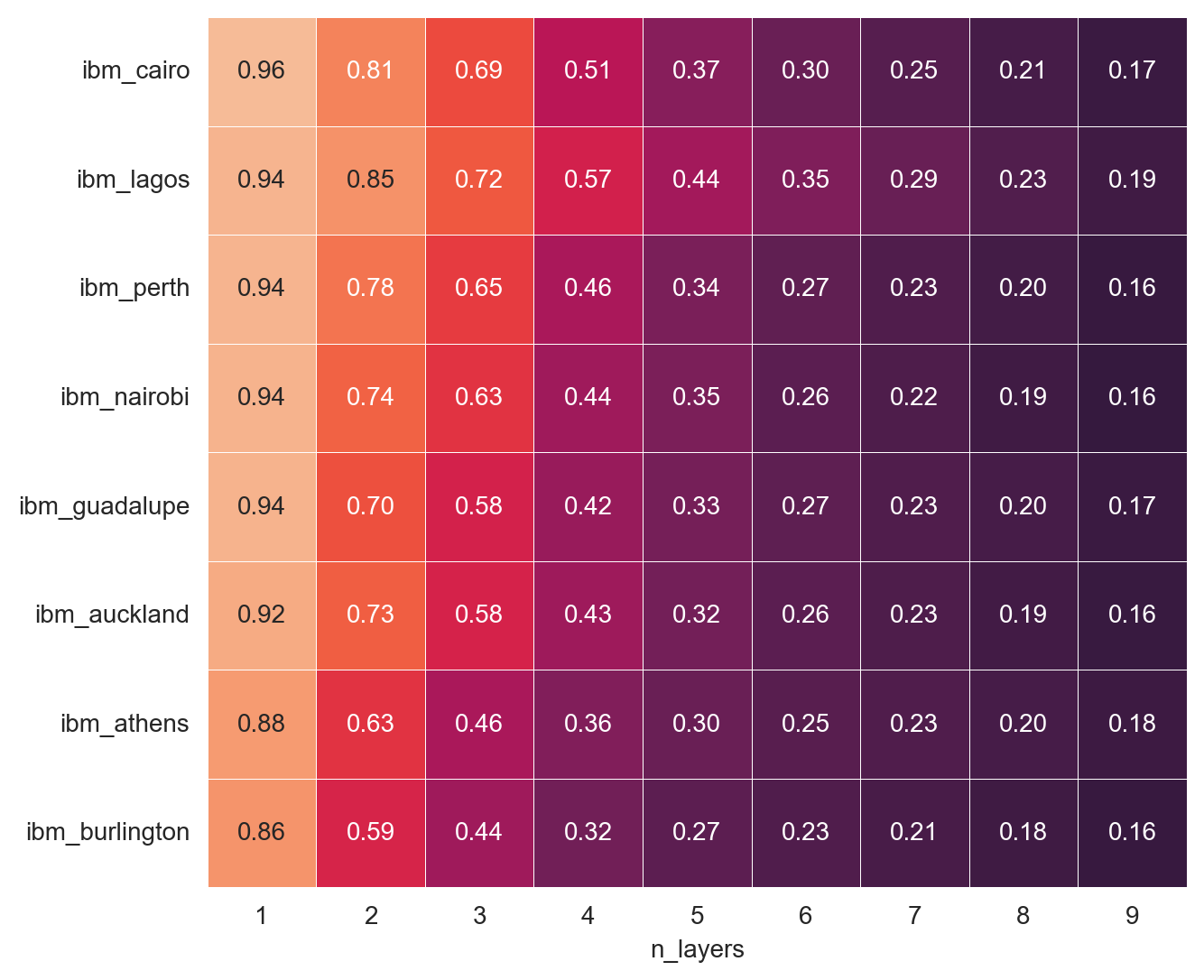}
    \caption{Benchmarking results on several IBM QPUs with 5 qubits. The chart is sorted by the H-Scores with \texttt{n\_layers}=1.}
    \label{fig: result 5 qubits}
\end{figure}

After our proof of concept, we proceeded to benchmark various IBM Quantum's QPUs using HamilToniQ. The results for 3, 4 and 5 qubits are presented in Fig.~\ref{fig: result 3 qubits}, Fig.~\ref{fig: result 4 qubits}, and Fig.~\ref{fig: result 5 qubits}, respectively. To conserve computing resources, ``fake providers'' were employed to simulate the performance of the actual QPUs. Detailed specifications of these QPUs are available on the IBM Quantum Cloud Server.

From the benchmarking results, we can compare the performance of various superconducting-circuit processors on the IBM Quantum Cloud Server. Specifically, in tasks involving the QAOA with a small number of qubits (from 3 to 5 qubits), \texttt{ibm\_cairo} and \texttt{ibm\_lagos} provide the most stable performance as the number of layers increases. Conversely, \texttt{ibm\_burlington} exhibits the worst performance in the benchmarking. With an increasing number of layers, the H-Score decreases due to the cumulative effect of two-qubit gate errors and each qubit's coherence and decoherence times. This observation directly reflects the design of the device's topology and the physical gate protocols.

In general, the H-scores tend to increase with the use of more qubits, and some H-Scores may even reach or exceed \(1\). This occurs because the gradient in the QAOA algorithm's parameter space becomes exponentially flatter as more qubits are used, which causes the optimization algorithms to become less efficient. Consequently, lower accuracy is obtained from an ideal simulator, leading to a diminished impact of noise on the accuracy, and occasionally even contributing ``positively'' to the outcomes.

Certain QPUs exhibit higher H-Scores than others when operating with a small number of layers, but this advantage diminishes faster than others with an increase in layer count. This phenomenon likely stems from the relatively shorter decoherence time of these processors, which introduces more inaccuracy with additional layers.

For example, when using 3 qubits, the QPU \texttt{ibm\_authen} initially outperforms \texttt{ibm\_perth} with 1 and 2 layers but then experiences a faster decline. The average decoherence times \(T_1\) and \(T_2\) for the qubits used in \texttt{ibm\_authen} are 56.49 \(\mu\)s and 99.62 \(\mu\)s, respectively, compared to 210.19 \(\mu\)s and 106.18 \(\mu\)s for those in \texttt{ibm\_perth}.

\subsection{Benchmarking the Topology of Quantum Devices}

When benchmarking QPUs with larger qubit numbers, the topology of the devices becomes increasingly important. This significance arises partly because the connectivity of each qubit influences performance, and partly because the physical limitations encountered when transpiling algorithms to real hardware can impact effectiveness. The influence of QPU topology has been extensively discussed in various research works focusing on superconducting circuits\cite{chamberland2020topological,sisodia2020comparison}, trapped ions\cite{pino2021demonstration}, and neutral atoms\cite{bluvstein2024logical}.

\begin{figure}[ht]
    \centering
    \includegraphics[width=1\linewidth]{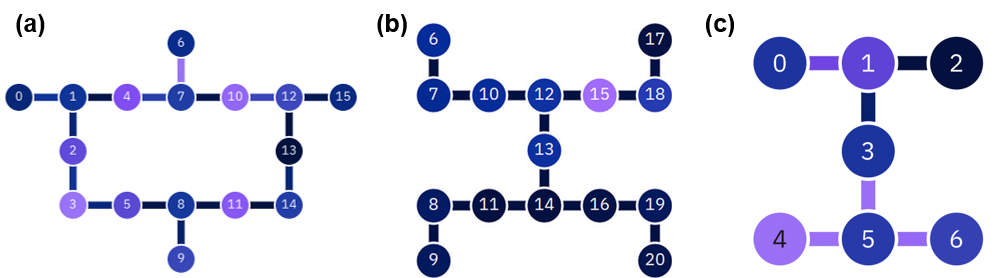}
    \caption{The different topologies of QPUs in the IBM Quantum Cloud Server are shown: (a) The Heavy Hexagonal Lattice, an example of which shown in our case is \texttt{ibm\_guadalupe}. (b) Connected two-line, an example shown in our case is \texttt{ibm\_auckland}. (c) I-Shape, examples shown in our case are \texttt{ibm\_perth} and \texttt{ibm\_lagos}.
}
    \label{fig: topology}
\end{figure}

\begin{figure}[ht]
    \centering
    \includegraphics[width=\linewidth]{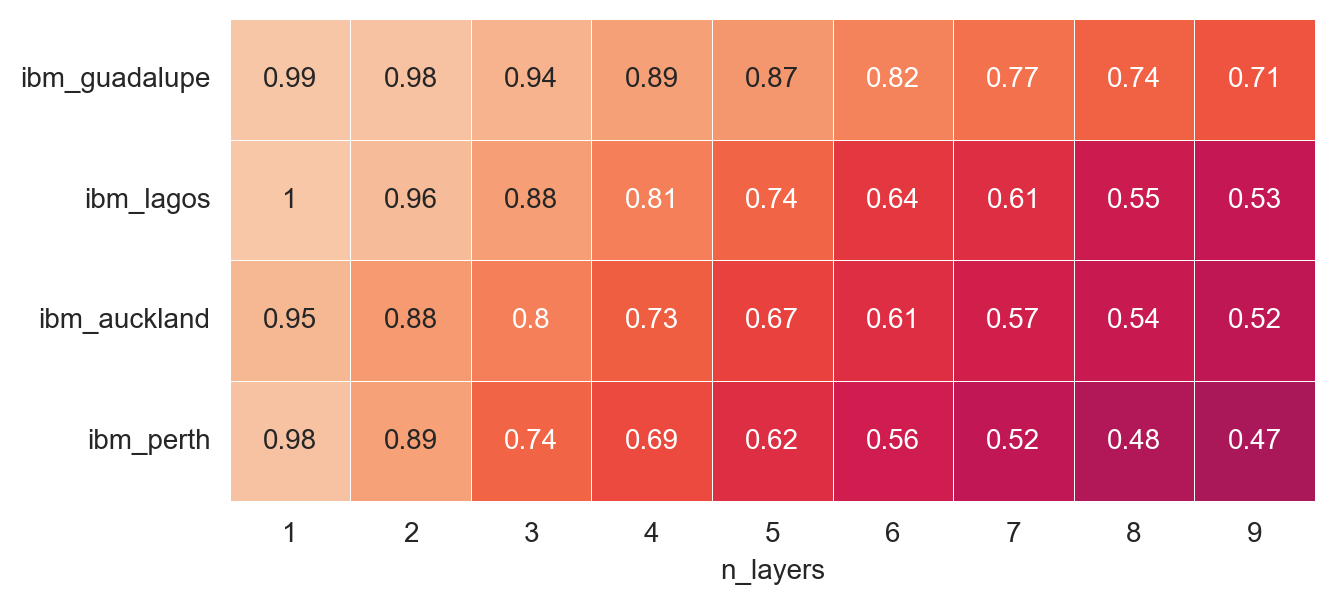}
    \caption{Benchmarking results for several IBM QPUs with 6 qubits, sorted by H-Scores at \texttt{n\_layers} = 9, focusing on QPU stability with large circuit depths. }
    \label{fig: result 6 qubits}
\end{figure}

In our benchmarking experiment, we explore three main different topologies: the heavy hexagonal lattice shown in Fig.~\ref{fig: topology}(a), the connected two-line topology shown in Fig.~\ref{fig: topology}(b), and the I-shape topology shown in Fig.~\ref{fig: topology}(c). In the benchmarking results shown in Fig.~\ref{fig: result 6 qubits}, we can see that \texttt{ibm\_guadalupe} exhibits the most stable performance as the number of layers increases, in comparison to other QPUs. Notably, the heavy hexagonal lattice demonstrates superior performance as the number of layers increases, outperforming other topologies.

\subsection{Benchmarking Error Suppression Protocols}

In this study, we utilize IBM Quantum Primitives for setting up QEM and DD within our QPU system. Specifically, we configured the system with \texttt{optimization\_level=3} for DD (to suppress decoherence errors \cite{niu2022effects}) and \texttt{resilience\_level=1} for QEM (employing the T-Rex protocol to suppress readout errors \cite{van2022model}).

\begin{figure}[ht]
    \centering
    \includegraphics[width=\linewidth]{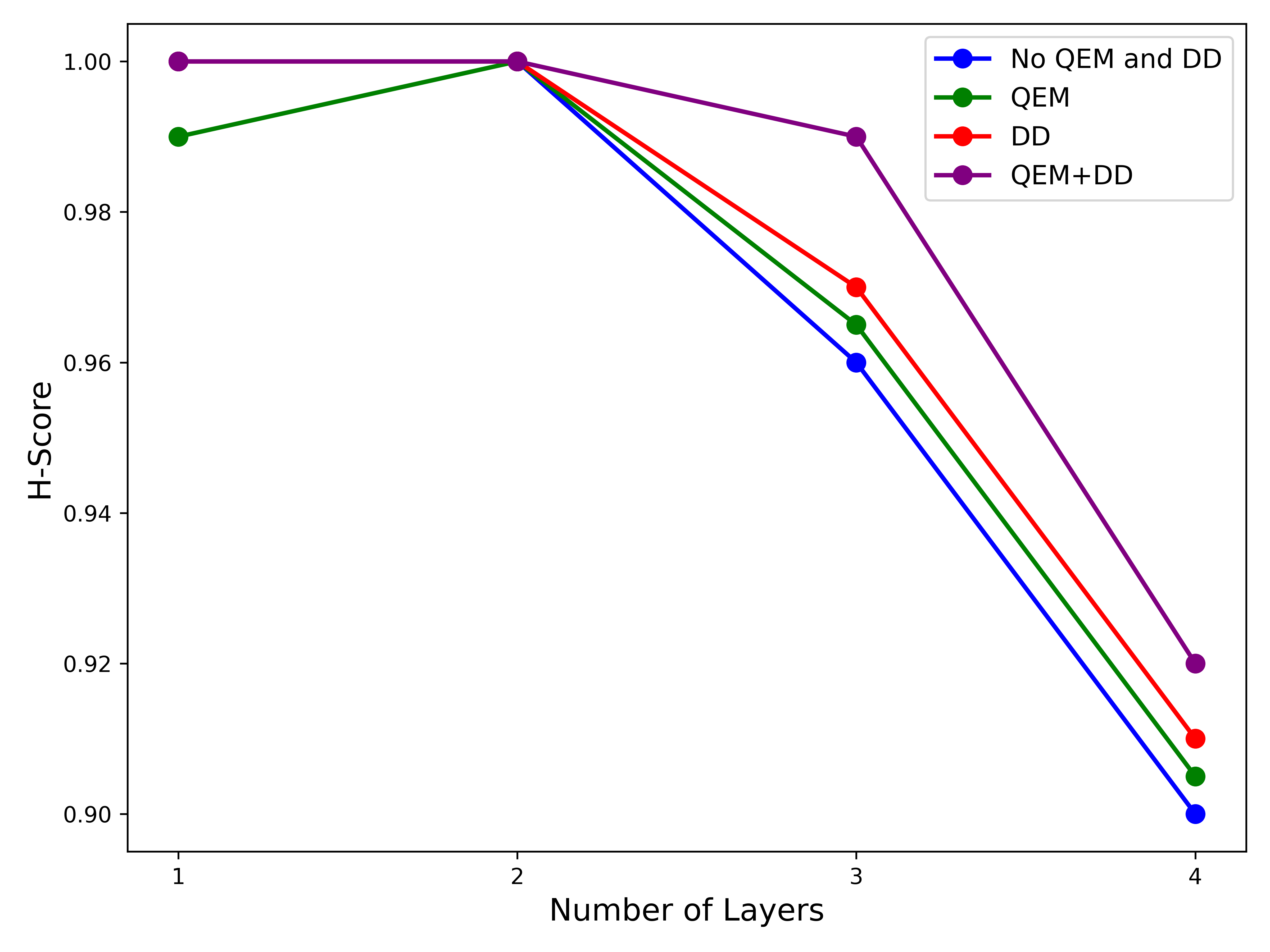}
    \caption{Comparison of H-score stability across one to four layers in systems employing QEM, DD, both, or neither. These results were obtained on \texttt{ibm\_guadalupe} with 6 qubits in use.}
    \label{fig: QEM_result}
\end{figure}

 As illustrated in Figure~\ref{fig: QEM_result}, this optimized compilation and quantum error mitigation strategy resulted in an enhanced H-Score during our benchmark tests. This benchmarking scheme is also applicable to hardware implementations, for example, recent Q-CTRL work on the IBM Quantum Cloud Server. Based on existing research concerning coherent control, we anticipate that the integration of Q-CTRL with IBM QPUs will further improve the H-Score \cite{mundada2023experimental}.


\subsection{Resource Management of Distributed Quantum System with HamilToniQ Benchmarking Toolkit}

HamilToniQ can also be used for distributed quantum system management with its benchmarking score record. For example, if we have 10 quantum processes in our quantum cloud server (shown in Fig.~\ref{fig: result 3 qubits}) and we want to utilize some of them to perform distributed quantum algorithms or distributed system sampling, we may need a ranking guide to help us choose the QPUs that are best for our use case to achieve optimal performance. However, evaluating QPUs based on their decoherence time of each qubit, readout error rate and quantum-gate-error rate is expensive due to the many variables that need to be considered. Hence, our benchmarking toolkit can provide a straightforward ranking to help us compile the algorithm we want to run on the most stable QPUs and achieve the best performance for our task.

\begin{figure}[htpb]
    \centering
    \includegraphics[width=\linewidth]{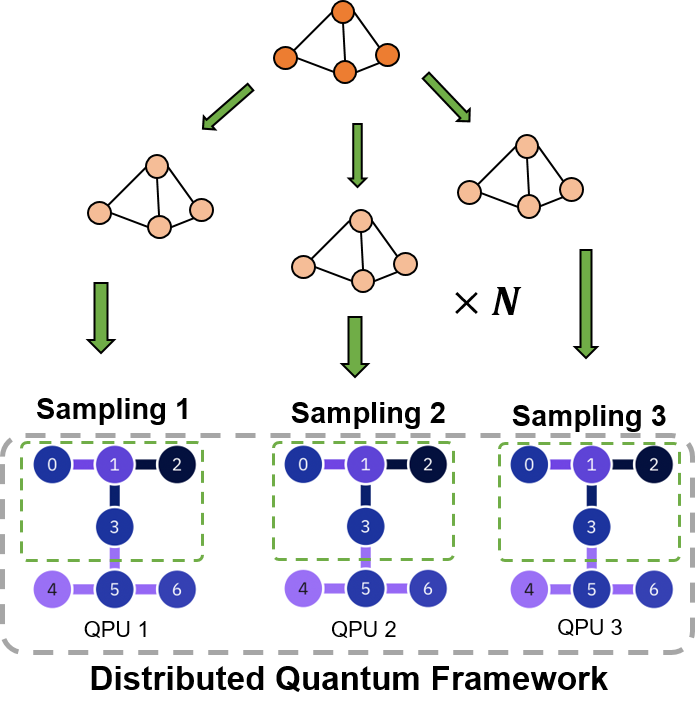}
    \caption{An example of the Distributed Quantum Framework, depicting three samplings of quantum states across QPUs. Each sampling captures unique quantum state interactions, contributing to a comprehensive ensemble essential for the distributed computing model.}
    \label{fig: dist-demo}
\end{figure}

\begin{figure}[htpb]
    \centering
    \includegraphics[width=0.9\linewidth]{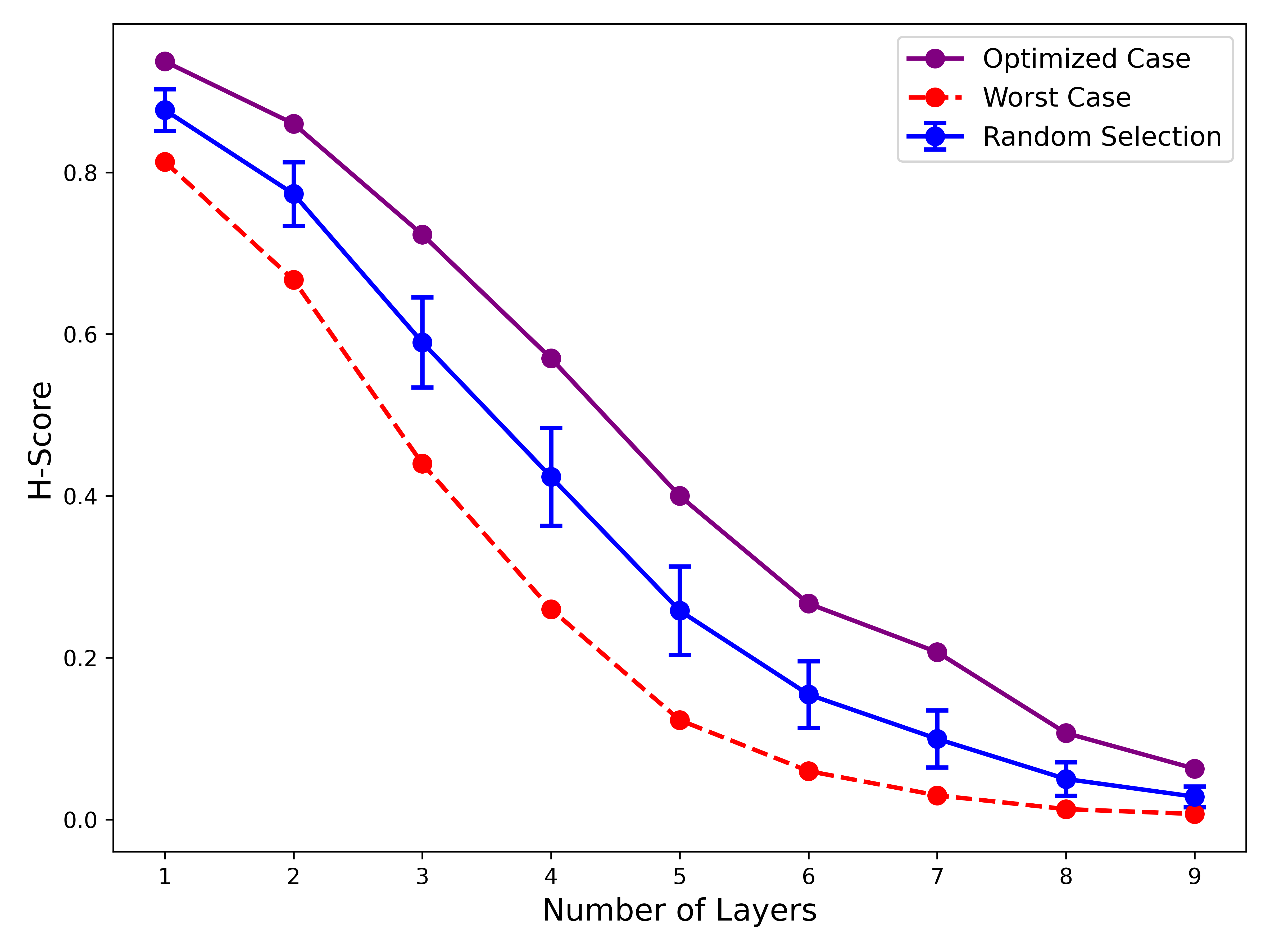}
    \caption{Comparison of H-Score stability across an optimized compilation strategy, benchmarked with HamilToniQ ranking, against random selection and the worst-case scenario.}
    \label{fig: dist-demo-result}
\end{figure}

In our demonstration, we selected a 3-qubit QAOA distributed sampling task as a proof of concept. We compared the worst-case scenario—where the compilation strategy was not optimized—with a strategy aligned with the HamilToniQ QPU ranking, as shown in Fig. 5, utilizing H-Score benchmarking. The results depicted in Fig.~\ref{fig: dist-demo-result} demonstrate that without optimized compilation (blue color), the performance is significantly lower than with optimized compilation (purple color). The random selection case (red color), represented by the blue curve with error bars, involves selecting 3 out of 10 QPUs randomly and was tested 200 times to derive the H-Score results. The worst-case and the optimized case exhibit an approximate 20\% improvement in H-Score. Thus, the The HamilToniQ toolkit not only provides a performance comparison among various QPUs and protocols but also offers a strategic framework for resource management in Quantum-HPC or distributed quantum systems \cite{cuomo2020towards, chen2024quantum}. Through this resource management scheme, we can enhance the capability of these systems in various fields such as portfolio optimization\cite{brandhofer2022benchmarking}, and routing problems\cite{harwood2021formulating, azad2022solving}.

\subsection{Robustness of the Benchmarking Results}

\begin{figure}[htpb]
    \centering
    \includegraphics[width=0.85\linewidth]{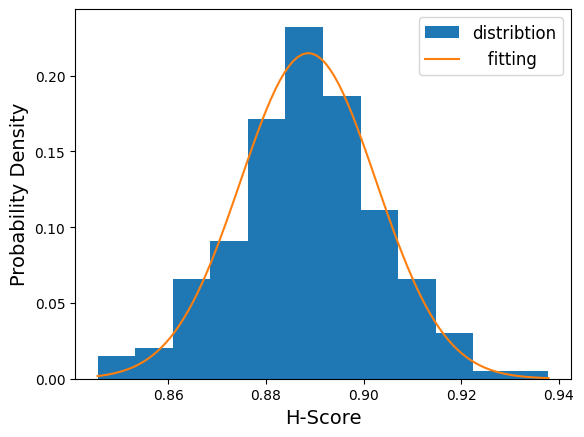}
    \caption{The distribution and its Gaussian fitting (orange curve) of the H-Scores (blue histogram) on \textit{ibm\_lagos} processor.}
    \label{fig: stability}
\end{figure}

If HamilToniQ is executed multiple times on the same processor with the same number of layers, the H-Scores of this QPU are not static but exhibit a distribution. This section delves into the analysis of such a distribution. As an example, the distribution of H-Scores from the \textit{ibm\_lagos} QPU is shown as a histogram in Fig.~\ref{fig: stability}, illustrating 200 H-Score measurements utilizing 3 qubits and 2 layers. A Gaussian fit is applied to this distribution, revealing a mean of $0.8887$ with $95\%$ confidence bounds of $0.8872$ and $0.8901$. The standard deviation from the fit is $0.0139$, with $95\%$ confidence bounds of $0.0124$ and $0.0153$. This relatively narrow standard deviation facilitates reliable performance comparisons from the initial run of HamilToniQ. Furthermore, the non-Gaussian PDF evaluation curve, shown in Fig.~\ref{fig: stability}, is transformed into a Gaussian distribution through statistical methods with multiple measurements (default is 200 times), enabling consistent benchmarking of performance across different scenarios and demonstrating the robustness of our benchmarking scheme.

\section{Conclusion}
In this study, we introduce HamilToniQ, a cutting-edge benchmarking toolkit for quantum systems. Specifically designed for application-oriented benchmark of QPUs through quantum optimization algorithms like QAOA, HamilToniQ utilizes the H-Score to evaluate and manage QPU performance. We give a comprehensive theoretical derivation of the H-Score along with an analysis of its robustness, validating its applicability with the QAOA algorithm. We also conducted a series of validations on different QPUs, examining aspects such as their topology and the effectiveness of error suppression protocols like QEM and DD. Furthermore, we demonstrate the potential of this toolkit for multi-QPU resource management within a Quantum-HPC hub or distributed quantum system, validating the efficacy of the H-Score for multi-QPU processing. hus, the HamilToniQ benchmarking scheme can assist these systems in handling the multi-QPU scheduling problem with low overhead, effectively solving large-scale problems such as portfolio optimization and vehicle routing. Furthermore, HamilToniQ marks a significant advancement by offering a scalable assessment framework that covers the entire quantum software stack. This framework enables empirical evaluations of quantum software tools, ensuring they are repeatable, comparable and transparent. Public access to our open-source Github repository is provided in Code Visibility. 


\section*{Acknowledgment}
The authors extend their gratitude to Nils Quetschlich for invaluable insights, which have been pivotal to the success of this research work. K.C. is grateful for the financial support from both the Turing Scheme for the Imperial Global Fellows Fund and the Imperial QuEST Seed Fund. Special acknowledgment goes to the IBM Quantum Researcher Programme and the research framework from QuantumPedia AI.

\section*{Code Visability}
\label{code_vis}
The source code for generating the dataset and figures presented in this research work is openly accessible at \url{https://github.com/FelixXu35/hamiltoniq}. During the development and testing of HamilToniQ, several Python libraries were essential. Specifically, Qiskit version 1.0.2, Qiskit-Aer version 0.13.3, Qiskit-Algorithm version 0.3.0, and Qiskit-IBM-Runtime version 0.21.2 were utilized.

\addcontentsline{toc}{chapter}{Bibliography}
\bibliographystyle{unsrt} 
\bibliography{references}

\end{document}